\documentstyle[epsf,prd,eqsecnum,aps]{revtex} 
\begin{document}
\draft
\title{Spherical scalar waves and gravity - red shift and backscattering}

\author{Edward Malec}
\address{Jagiellonian University,
Institute of Physics, {3}0-59 Krak\'ow, Reymonta 4, Poland}
 \author{Niall \'O Murchadha}
\address{Physics Department, University College, Cork, Ireland}
\author{Tadeusz Chmaj}
\address{Institute of Nuclear Physics, Krak\'ow, Radzikowskiego 152, Poland} 
\maketitle
\begin{abstract}
This article investigates the interaction of a spherically symmetric
 massless scalar field with a
strong gravitational field. It focuses on the propagation of
 waves in regions outside any horizons. The two factors acting on the waves
 can be identified as a redshift and a backscattering.
The  influence of backscattering on the intensity
 of the outgoing radiation is studied and
rigorous quantitative upper bounds obtained. These show 
that the total flux may be decreased if
the sources are placed in a region adjoining an apparent horizon.
 Backscattering can be neglected in
the case
$2m_0 /R<< 1$, that is when the emitter is located at 
a distance from a black hole much larger
than the Schwarzschild radius. This backscattering may have
 noticeable astrophysical consequences.
\end{abstract}

\section{ Introduction.}

A black hole   can   be investigated by observing the gravitational
radiation     or the electromagnetic radiation emitted by a medium surrounding
it. The two major effects that act on the radiation are redshift
 and  backscatter. While the
redshift is well understood, little attention has been focused to
 date on backscattering, a
phenomenon that prevents radiation from being transmitted exclusively 
  along  null cones.
Backscattering can   influence   observations, since sharp
 and strong initial impulses  can be
weakened and dispersed.  Electromagnetic or gravitational radiation, for 
instance, reaching an
asymptotic observer,  can be weaker than expected.

In this paper we  assess analytically this effect by considering as  
a toy model a spherically
symmetric  massless scalar field.   This simplifying assumption should not
 restrict  seriously the
validity of   our conclusions.  We study the evolution of pulses
 of radiation which are initially
purely outgoing and  emitted  from a region close to a spherical black hole.
 Rigorous upper bounds
on the magnitude of the backscattering effect are obtained; these estimates,
 while they   break down
close to the horizon, allow  one to  recognise the regions  in which the
 backscattering may be
neglected.  We  estimate how significant this effect would be 
in the case of neutron stars or black
holes. The analytic estimates that are obtained here  can be 
used in numerical gravity in order to
assess numerical codes. They can be used to test both the accuracy 
and the long-time stability of
the numerical models.  The final section    outlines
    possible astrophysical consequences of
backscattering.

Our results give  the first, we believe,  quantitative measure 
of this strong field effect.
 It is  worthwhile pointing out that both our approach
 and the physical problem
dealt with are   different from that discussed in
 the existing literature on backscattering,
which  concentrates on the  analysis of stationary
 radiation on a fixed geometric background
\cite{bachelot}.

\section{Spherically symmetric massless scalar fields in Minkowski space}

Let us consider Minkowski space with a spherically
 symmetric coordinate system and
 a metric $g_{\mu \nu} = (-1, +1, R^2, R^2\sin^2\theta)$. 
 Let  $\phi(R,t)$ be  a
spherical field which satisfies the simplest massless wave equation
\begin{equation} \nabla^{\mu}\nabla_{\mu} \phi = 0\,. \label{2.1}
 \end{equation}
 The general solution (except at the origin) to Eq.(\ref{2.1})  is
\begin{equation} \phi = {f(R + t) + g(R - t) \over R}\,, 
\label{2.2} 
\end{equation}
where $f$ and $g$ are
 essentially
arbitrary functions of one variable. 
Therefore the general solution
 can be regarded as a superposition of an ingoing null
 field $f$ and an outgoing null field $g$.
Associated with the field $\phi$ will be a stress-energy tensor
 $T_{\mu \nu}$, an object of
some importance,  as we want to couple the scalar field to gravity.'
 {\it A priori},
 it is required only  that $T_{\mu \nu}$ be a symmetric tensor
 that satisfies the conservation
equation
$T^{\mu}_{\nu;\mu} = 0$. The standard choice is
\begin{equation}
 T_{\mu \nu} =
\nabla_{\mu}\phi\nabla_{\nu}\phi - {1 \over 2}
 g_{\mu \nu} (\nabla\phi)^2\,.
\label{2.3}
 \end{equation}
This  tensor  will be subsequently used, 
but the reader should be aware that it
is not necessarily unique.

Given the way that the field runs along the ingoing and outgoing null
directions, it  is natural to consider the two ingoing
 and outgoing null derivatives of $\phi$
\begin{eqnarray} U &= (\partial_R + \partial_t)\phi &=
 {2f' \over R} - {\phi \over R} \label{2.4}\\
V &= (\partial_R - \partial_t)\phi &= {2g' \over R} -
 {\phi \over R}~. \label{2.5} \end{eqnarray}
When written  in terms of  $U$ and $V$, the 
stress-energy tensor  take  a form
\begin{eqnarray}
 T_{00} &={1 \over 4}(U^2 + V^2) &= \rho\label{2.6}\\
 T_{0R}&= {1 \over 4}(U^2 - V^2)
 &= -J\label{2.7}\\ T_{RR}
&= {1 \over 4}(U ^2 + V^2)\\ T_{\theta\theta}
&= - {1 \over 2}R^2 UV\\
T_{\phi \phi}&= -{1 \over 2}R^2 \sin^2\theta UV~.
\label{2.10}
\end{eqnarray}
From Eqs.(\ref{2.6}) and (\ref{2.7})  one can be tempted to  interpret
$ V^2$       and $U^2$ as, respectively, outgoing and
  ingoing  components of a radial     flux.
There is, however,   a discrepancy with what can be'
 inferred from (. \ref{2.2}); from
that follows, for instance,  that field is purely outgoing 
if $f \equiv 0$.
When   returning to Eqs.(\ref{2.4}) and (\ref{2.5}), it is clear that   then    $U$ does not
vanish identically.   As a consequence, if the field is purely
 outgoing, the energy-momentum has an
inward part which only vanishes asymptotically so that the 
energy-momentum is timelike even though
the field is null.

 There do exist objects which more naturally reflect the splitting given by
Eq.(\ref{2.2}) into ingoing and outgoing modes. Let us define
\begin{eqnarray} h_+(R,t) &= {1 \over 2} (\partial_R - \partial_t) 
(R\phi) &= g'
= {RV \over 2} + {\phi \over 2}\label{2.11}\\ h_-(R,t) &=
 {1 \over 2} (\partial_R + \partial_t)
(R\phi) &= f' = {RU \over 2} + {\phi \over 2}~.\label{2.12} \end{eqnarray}
  Extending  $R$ to cover the whole real line and defining $h(R,t)$ via
\begin{eqnarray} h(R,t) &= h_+(R,t),& R>0\\ h(R,t) &= h_-(R,t),& R \le 0~,
\end{eqnarray}
allows one to write  the field equation (\ref{2.1})  compactly  as
\begin{equation} (\partial_R + \partial_t)h(R,t) = 0~.
\end{equation}
By comparison, when  writing earlier the field equation in terms of $U$ and $V$ we
get a pair of coupled equations. There exists another stress-energy
 tensor, more naturally related
to $h_{\pm}$, of the form \begin{eqnarray} T'_{00} &=& {h^2_+ +
 h^2_- \over R^2} \\ T'_{0R} &=&
{h^2_- - h^2_+ \over
 R^2} \\ T'_{RR} &=& T'_{00}\\ T'_{\theta\theta} &=& T'_{\phi \phi} = 0~. 
\end{eqnarray}
This also satisfies $T'^{\mu}_{\nu ; \mu} = 0$ in flat space. Unfortunately,
 it is not clear how to generalize  that object  to curved space 
so we will restrict our attention
to the $T^{\mu\nu}$ as defined by Eqs.(\ref{2.6} - \ref{2.10}). 
 While the field
equation is best expressed in terms of $h(R,t)$,
 as will be shown later, the  conserved
quantities  are natural in terms of $U$ and $V$.

 The total energy inside a sphere of radius $R$ at a fixed time $t$ reads
\begin{equation} M(R,t) = 4\pi\int^R_0
r^2 \rho dr = 4\pi \int^R_0 r^2 T_{00}dr = 
\pi\int^R_0 r^2 (U^2 + V^2)dr~.\label{2.16}
\end{equation}
 It is now easy to compute various transport  equations for the 
 energy $M$
\begin{eqnarray} \partial_RM(R,t) &= 4\pi R^2\rho &= \pi R^2(U^2 + 
V^2)\\ \partial_tM(R,t)
&= -4\pi R^2 J &= \pi R^2(U^2 - V^2)\\ (\partial_R + \partial_t)M(R,t) &= 
2\pi R^2
U^2\label{2.19}\\ (\partial_t - \partial_R)M(R,t) &= -2\pi R^2 V^2
\label{2.20}~.
\end{eqnarray}
 It is clear from Eqs.(\ref{2.19}) and (\ref{2.20}) that $V^2$ and $U^2$
 can be interpreted as  the
energy fluxes in the outgoing future and ingoing future null directions 
respectively.

\section{Spherically symmetric massless scalar fields in
 Schwarzschild spacetime}

Let us now consider a spherically symmetric massless scalar 
field minimally coupled
  to the Schwarzschild spacetime with the standard external metric
\begin{equation} ds^2 = -\left(1 - {2m_0 \over R}\right)dt^2 + \left(1 -
{2m_0 \over R}\right)^{-1}dR^2 + R^2d\Omega^2~,
\end{equation}
where $m_0$ is a positive constant, the Schwarzschild mass. 
While  there is no
explicit solution  akin to Eq.(\ref{2.2}), much of the analysis of
 Section 2 carries over. Define
a `radiation amplitude'
$h(s,t)$, $s\epsilon (-\infty , \infty )$, as follows
 \begin{eqnarray}
 h_+(R, t) &= h(R,
t)&={1\over 2} (-{1\over \gamma }\partial_0 +\partial_R ) (R\phi ) 
\nonumber\\ h_-(R, t) &= h(-R,
t)&={1\over 2} ({1\over \gamma }\partial_0 +\partial_R ) (R\phi )~. 
\label{3.1}
\end{eqnarray}
where $\gamma = 1 - 2m_0/R$. Given $h(R)$   the field $\phi$ 
can be recovered by a simple
integration
\begin{equation} \hat h= -{1\over 2R}\int_{R}^{\infty}dr[h_+(r) +
h_-(r)]={1\over 2}\phi ~.
\label {3.2}
\end{equation}
 The scalar field
equation, $\nabla_{\mu }\partial^{\mu } \phi =0 $, Eq.(\ref{2.1}),
 can be written as a single
first order equation on a `symmetrized' domain $-\infty \le R\le \infty $,
\begin{equation} (\partial_0 + \gamma \partial_R)h = (h -\hat h ) 
{-2m_0R \over|R|^3} ~.
\label {3.3}
\end{equation}
 It is useful to change coordinates and introduce
 the Regge-Wheeler coordinate \cite{MTW} as a
new   radial variable
\begin{equation} r^* = R+2m_0 \ln ({R\over 2m_0}-1).
\label {3.4}
 \end{equation}
Using this, equation (\ref {3.3}) reads
\begin{equation} (\partial_0 + \partial_{r^*})h =
 (h -\hat h ) {-2m_0 R \over |R|^{3}}~.
\label {3.5}
 \end{equation}
From Eq.(\ref{3.4})  follows that
$dR/dr^* = 1 - 2m_0/R$ and this allows us to write Eq.(\ref{3.5}) as
\begin{equation} (\partial_0 + \partial_{r^*})[(1 - {2m_0 \over R})h] =
 {2m_0 R \over |R|^3} [(1 -
{2m_0 \over R})\hat h]~.
\label{3.6}
 \end{equation}
 This can be solved by
\begin{equation}
(1 - {2m_0 \over R})h(r^* , t) = h_0(r^* - t) +
\int_{(r^* - t, 0)} ^{(r^* , t)} {2m_0 R \over |R|^3}[(1 - 
{2m_0 \over R})\hat h] dv~,
\label{3.7}
\end{equation}
 where the integration in Eq.(\ref{3.7})  extends 
 along the `outgoing' null ray from a point
on the initial slice at $(r^* - t, 0)$ to $(r^* , t)$.
 (Of course  in the negative $R$
region, where we are dealing with $h_-$, the 
integration  contour follows along   the physical
ingoing null ray.) It is clear that $\hat h$   
mixes the outgoing with the ingoing modes and
thus is responsible for the backscattering. 
The leading term in Eq.(\ref{3.7}) is just the standard
gravitational redshift. One gets from  Eq.(\ref{3.7}) at $t = 0$
\begin{equation}
 h_0(x) = \left(1 - {2m_0 \over R(x)}\right)h(x, 0)~,
 \end{equation}
and ignoring the second term in Eq.(\ref{3.7})  we  arrive at
\begin{equation} h(r^*(0) + \tau, \tau) =
\left(1 - {2m_0 \over R(r^*(0))}\right)
 \left(1 - {2m_0 \over R(r^*(0) + \tau)}\right)^{-1} h(r^*(0), 0)~,
 \label{3.10}
\end{equation}
when $\tau$ becomes large.

As in the Minkowski space-time case  
the stress-energy Eq.(\ref{2.3}) naturally  expressess
via $U$ and $V$ defined below
\begin{eqnarray} U &=
(\gamma^{1 \over 2}\partial_R + \gamma^{-{1 \over 2}}\partial_t)\phi \\ V &=
(\gamma^{1 \over 2}\partial_R - \gamma^{-{1 \over 2}}\partial_t)\phi ~.
\end{eqnarray}
The relevant components of the stress-energy tensor read
\begin{eqnarray}
 \rho &= -T^0_0 &= {1 \over 4}(U^2 + V^2)\\ J &= -T_{0R} &=
{1 \over 4}(V^2 - U^2)\\ T &= T^R_R &= {1 \over 4}(U^2 + V^2)~.
\end{eqnarray}
 The `mass function' - the energy associated
 with a ball of radius R - is given by
\begin{equation}
 m(R,t) = 4 \pi
\int_0^R r^2 \rho dr~.
\label{3.16}
\end{equation}
 This is the analogue of the
Misner-Sharp-Hawking-{\it et al} mass.
 Differentiation of Eq.(\ref{3.16}) yields
\begin{eqnarray}
 \partial_Rm(R,t) &= 4\pi R^2\rho &= \pi R^2(U^2 + V^2)
\label{3.17}\\
\partial_tm(R,t) &= -4\pi R^2 J &= \pi \gamma R^2(U^2 - V^2)~.
\label{3.18}
\end{eqnarray}
  The derivatives of $m$ along the outgoing and ingoing null
 directions can be written,
 employing Eqs.(\ref{3.17}) and (\ref{3.18}), as
\begin{eqnarray}
(\gamma\partial_R + \partial_t)m(R,t) &= 2\pi \gamma R^2 U^2
\label{3.19}\\ (\partial_t -
\gamma\partial_R)m(R,t) &= -2\pi \gamma R^2 V^2.
\label{3.20}
\end{eqnarray}
Let us comment on the derivation of  Eq.(\ref{3.18}),
  since it is not obvious and  since
that  would help to understand  why  $m(R,t)$  
is chosen as above, in a non-canonical
way.  There is a timelike Killing vector $\xi^{\mu} = (1,0,0,0)$ which
 satisfies $\xi_{(\mu;\nu)}$.
This implies, together with the stress-energy
 conservation law $T^{\mu}_{\nu;\mu} = 0$,
$(T^{\mu}_{\nu}\xi^{\nu})_{;\mu} = 0$. 
The  vectorial divergence  can be written  as
$(\sqrt{-g}T^{\mu}_{\nu}\xi^{\nu})_{,\mu} = 0$;  this  leads to
\begin{equation} -
(\sqrt{-g}T^0_0)_{,0} = (\sqrt{-g}T^a_0)_{,a}~.
\label{3.24}
\end{equation}
The integral of the left hand side of Eq.(\ref{3.24})
 over a sphere is clearly $\partial m/\partial
t$ from Eq.(\ref{3.16}). The integral of the right hand
 side is just a flat space total divergence
which becomes a surface integral. From this   Eq.(\ref{3.18})
 immediately  follows.

\section{The spherically
symmetric Einstein - massless scalar field system}

In this  section we will study the full Einstein-scalar 
field system, including the backreaction
that the matter exerts on the geometry.
 The polar gauge condition $tr K = K_r^r$  is our slicing
condition. This guarantees that the normal to the slices
 is along the lines of constant $R$. Its
major  attraction  is that to a large degree 
it eliminates the extrinsic curvature from the field
equations. This allows one to express 
 the metric explicitly in terms of the matter-field. The
great disadvantage is that one cannot cope with horizons 
in the polar gauge, thus  the latter is not
suitable for the descriptions of late 
stages of the collapse but it is useful in the exterior
zone that we wish to investigate.
  The aforementioned  choice of gauge is equivalent to choosing  a
diagonal line element
\begin{equation} ds^2 = - \beta (R, t)\gamma (R, t)dt^2 
+ {\beta (R, t)\over \gamma (R, t)} dR^2 +
R^2(r, t) d\Omega^2~,
 \label{4.1}
 \end{equation}
where $t$ is a time coordinate, $R$ is a radial
coordinate that coincides with the 
areal radius and $d\Omega^2 = d\theta^2 + \sin^2\theta d\phi^2$
is the line element on the unit sphere, $0\le \phi < 2\pi $ 
and $0\le \theta \le \pi $. At spatial
infinity (of an asymptotically flat spacetime)
 the metric coefficient $g_{RR}$ tends to $+1$ and
$g_{tt}$ tends to $-1$.  Thus   $\beta$ and $\gamma$ go to $+1$ at infinity.
We adopt the standard convention that Greek 
letters range from 0 to 3 for spacetime objects while
Latin indices range from 1 to 3 for spatial objects. 
The Einstein equations $G_{\mu\nu} = R_{\mu
\nu }-g_{\mu \nu }R/2 = 8\pi T_{\mu \nu }$ 
can be written as a $1+{3}$ system of initial
constraints and evolution equations \cite{Wald}.
 Let    $\Sigma_t $ be a foliation of the spacetime by
Cauchy hypersurfaces, with $g_{ij}$ the intrinsic metric
 and $K_{ij}$ the extrinsic curvature. In
the  convention of Wald \cite{Wald},
$2N K_{ij}= + \partial_tg_{ij}$ and $trK =g^{ij}K_{ij}$.

 As before, the scalar field is massless and
 spherically symmetric, with the stress-energy given by
Eq.(\ref{2.3}). The matter energy density is $\rho =-T_0^0 $ 
and the matter current density is $J =
-T_{0R}/\beta$. As  above, we
 define a `radiation amplitude' $h(s)$, $s \epsilon (-\infty , \infty
)$,  as follows
\begin{eqnarray}
h_+(R,t) &= h(R, t) &={1\over 2}
 (-{1\over \gamma }\partial_0 +\partial_R ) (R\phi )\\
h_-(R,t) &= h(-R, t)&={1\over 2} 
({1\over \gamma }\partial_0 +\partial_R ) (R\phi ).
\label{4.2}
\end{eqnarray}

One can show \cite{Malec1997} that
\begin{equation}
\beta (R) = e^{-8\pi (\int_R^{\infty }+
\int_{-\infty}^{-R }) {dr\over r} ( h -\hat h)^2 }~,
\label {4.3}
\end{equation}
where
\begin{equation}
\hat h= -{1\over 2R}\int_{R}^{\infty}dr[h_+(r) + h_-(r)]={1\over 2}\phi~.
\label {4.4}
\end{equation}
  $\gamma $  can be expressed in the following useful form 
\cite{Malec1997} 
\begin{equation}
\gamma (R)={1\over R}\int_0^R\beta dr. \label {4.5}
 \end{equation}

In spherically symmetric gravity one knows that 
the gravitational field is kinematical,
any real dynamics is in the matter field. 
Therefore  the metric can be expressed in terms of the
matter. In the polar gauge this can be done explicitly
 (see Eqs. (\ref{4.3} - \ref{4.5})).

The scalar field equation
$\nabla_{\mu }\partial^{\mu } \phi =0 $
can be written as a single first
order equation on a `symmetrized' domain $-\infty \le R\le \infty $,
\begin{equation}
(\partial_0 +\gamma \partial_R)h = (h -\hat h ) {\gamma -\beta \over R}~.
\label {4.6}
\end{equation}
This equation, together with definitions of $h,
 \hat h , \beta $, and $ \gamma $,
encodes the all information carried by the
 Einstein equations coupled to the scalar
field. Notice that $\int_{-\infty }^{\infty }dr h(r)=
 \int_0^{\infty }dr[ h(r)+h(-r)]=
\int_ 0^{\infty }dr \partial_r\tilde \phi =0$,
 where $\tilde \phi = R\phi$; therefore initial data,
$h_0$, of compact support must satisfy the condition
\begin{equation}
\int_ {-\infty }^{\infty }dr h_0 =0.
\label {4.7}
\end{equation}
It has been shown  shown, using
relations between metric functions and their symmetry properties,
 that if (\ref {4.7})
holds true then $\int_{-\infty }^{\infty }dr h(r, t_0)=0$ 
in the existence interval of a solution.
The local existence of a Cauchy solution for the above system and a global
existence in an external region bounded from the
 interior by a null cone have also
been shown (\cite{Malec1996b}, \cite{Malec1997}).

An alternative representation of $\gamma $ that will be useful is
\begin{equation}
\gamma (R) = \bigl( 1- {2m_0\over |R|} +{ 2 m_{ext}(R)\over R}\bigr) \beta (R)
\label {4.8}
\end{equation}
where $m_0$ is the asymptotic mass and $m_{ext}$ is a contribution to
the asymptotic mass coming from the exterior of a sphere of a radius $|R|$,
\begin{equation}
m_{ext}(R) =4\pi\int_{|R|}^{\infty }
{\gamma \over \beta } \Bigl( [h(r)-\hat h ]^2+
[h(-r)-\hat h ]^2\Bigr) dr~.
\label {4.9}
\end{equation}
Paralleling Eq.(\ref{3.19}) we can write
\begin{eqnarray}
(\partial_0 + \gamma\partial_R)m_{ext}(R) &= -8\pi\gamma^2 (h_- - \hat h)^2
\label{4.10}\\
(\partial_0 - \gamma\partial_R)m_{ext}(R) &= 8\pi\gamma^2 (h_+ - \hat h)^2
\label{4.11}
\end{eqnarray}
Let us stress that many of the equations written
 down in this section take on a quite special form
  in the polar gauge. In particular, the object
 defined in Eq.(\ref{4.9}) is the Hawking
mass (or mass function) which in a general gauge
 depends on the matter current as well
as on the energy-density. In the polar gauge it 
simplifies,  since the two null expansions are equal
 and the term which depends on the current drops out. 
Let us point out again that $h(R)$, $R>0$,
represents outgoing radiation while $h(-|R|)$ is an ingoing scalar wave.

We can substitute Eq.(\ref{4.8}) into Eq.(\ref{4.6})
 and rewrite the scalar field equation as
\begin{equation}
(\partial_0 +\gamma \partial_R)h = (h -\hat h ) {-2m(R)R\beta \over |R|^3}~,
\label {4.12}
\end{equation}
where $m(R) = m_0 - m_{ext}(R)$ is the Hawking mass at a radius $R$. This  is
  similar to Eq.(\ref{3.6}) but now backreaction is taken into account. 
 Therefore our
interpetation of Eq.(\ref{3.6}) as giving a `red-shift' due
 to the $h$ term on the right-hand-side
(determined by the local mass function, $m(R)$,
 rather than the Schwarzschild mass,
$m_0$) and a `backscattering' due to the $\hat h$ term
 should remain valid in some
appropriate limit.

In going from Eq.(\ref{3.5}) to Eq.(\ref{3.6})
 we found what was essentially an integration
function, $(1 - 2m_0/R)$, which allowed us eliminate 
the term in $h$ on  the right-hand-side of
Eq.(\ref{3.5}). It turns out that   this trick 
 can be repeated with Eq.(\ref{4.12}).

Let us define
\begin{equation}
\ln\left[1 - {2\tilde{m}(R) \over R}\right] = -\int_R^{\infty}{2m(r) dr
\over r^2(1 - 2m(r)/r)}~,
\label{4.13}
 \end{equation}
where the integral is taken along an outgoing null ray. 
 From Eq.(\ref{4.11}) follows
that the mass function monotonically increases along an outgoing null ray and   also
that $m(R) \rightarrow m_{BR_0}$
(the `local Bondi mass'   that can be assigned to a future null cone
starting from $R_0$).  Consider a null ray which starts at $R_0$; then
  $m(R_0) \le m(r) \le m_{BR_0}$. This immediately gives
\begin{equation}
{2m(R_0) \over r^2(1 - 2m(R_0)/r)} \le {2m(r) \over r^2(1 - 2m(r)/r)} \le
{2m_{BR_0} \over r^2(1 - 2m_{BR_0}/r)}~.
\end{equation}
Integration of this equation along the null cone  yields
\begin{equation}
1 - {2m(R_0) \over R} \ge 1 - {2\tilde{m}(R) \over R} \ge 
1 - {2m_{BR_0} \over R}~.
\label{4.14}
 \end{equation}
Since  $m(R_0) \rightarrow m_{BR_0}$ as $R_0 \rightarrow \infty$ we 
 conclude that
 $\tilde{m}(R) \rightarrow m_{BR_0}$.  $m(R)$ is
 positive along an outgoing null cone hence it is
clear that $1 - 2\tilde{m}(R)/R$, which is going to be
 our redshift factor, is monotonically
increasing. Note that there is  no guarantee that $ 1 - 2m(R)/R$ is monotonic.

Using Eq(\ref{4.13}) and Eq.(\ref{4.8}), it is easy to
 rewrite Eq.(\ref{4.12}) as
\begin{equation}
(\partial_0 + \gamma \partial_R)(1 - {2\tilde{m}(R) \over R})h = \hat h (1 -
{2\tilde{m}(R) \over R}){2m(R)R\beta \over |R|^3}~.
\label{4.15}
\end{equation}
 or as a pair of equations on the half-line, $R \ge 0$,
 \begin{eqnarray}
(\partial_0 + \gamma \partial_R)(1 - {2\tilde{m}(R) \over R})h_+ &=
\hat h (1 - {2\tilde{m}(R) \over R}){2m(R)\beta \over R^2} \label{4.16prim} \\
(\partial_0 - \gamma \partial_R)(1 - {2\tilde{m}(R) \over R})h_- &=
 -\hat h (1 - {2\tilde{m}(R) \over R}){2m(R)\beta \over R^2}~;
\label{4.16}
\end{eqnarray}
similar to equation (\ref{3.6}). Reasoning as in Section III
 leads to  a clean separation between
redshift and backscattering.

\section{weak field - redshift and backscatter}

We can define the whole problem in the exterior domain, independent of whatever
is happening in the interior. Already  $\hat h$, Eq.(\ref{4.4}),
is defined in a proper form. Given (\ref{4.7}), $\hat h$ can be written as
\begin{equation}
\hat h = -{1 \over 2R}\left[\int_{-\infty}^{-R} + 
\int_{+R}^{\infty}\right] dr h(r) =
{1 \over 2}\phi~.
\label{5.1}
\end{equation}
It is clear, from Eq.(\ref{4.3}), that $\beta(R)$ 
is only dependent on external
quantities. The  calculation of  $\gamma(R)$, 
 requires the use Eq.(\ref{4.5}), together with the
fact that, at infinity, $\gamma$ behaves like $1 - 2m_0/R$ 
while $\beta$ goes quickly to 1. This
allows us to write
\begin{equation}
\gamma(R) = 1 - {2m_0 \over R} + {1 \over R} \int_R^{\infty}[1 - \beta (r)]dr~;
\label{5.2}
\end{equation}
thus there are two equivalent expressions for $m_{ext}(R)$, Eq.(\ref{4.9}) and
\begin{equation}
m_{ext}(R) = {1 \over 2}\int_R^{\infty}[1 - \beta (r)] dr~.
\label{5.3}
 \end{equation}

 The form of Eq.(\ref{4.9})  inspires one to write Eq.(\ref{4.3}) as
\begin{equation}
\beta (R) = e^{-8\pi \left(\int_R^{\infty }+\int_{-R}^{-\infty }\right)
{\beta \over \gamma r} {\gamma \over \beta}( h -\hat h)^2 dr }~.
\label {5.4}
\end{equation}
    Eq.(\ref{4.8})  gives the inequality
\begin{equation}
{\beta \over \gamma r} = {1 \over r - 2m(r)} \le {1 \over r - 2m_0}~.
\label{5.5}
\end{equation}
The factor $\beta/\gamma r$ can be taken  out of
 the integral in Eq.(\ref{5.4})
and replaced   with its value at $r = R$.
 The remainder is then essentially Eq.(\ref{4.9}).
Select an $\epsilon$ and an $R_A$ such that 
simultaneously $m(R_A) \approx m_0$,
$m_{ext}(R_A)/m_0 <\epsilon$ and   $R_A > 
(1+\epsilon )  2m_0$. If $R > R_A $ then
\begin{equation}
1 \ge \beta(R) \ge \beta(R_A) \ge e^{-2{m_{ext}(R_A) 
\over R_A - 2m_0}} \ge e^{-{m_{ext}(R_A)
\over \epsilon m_0}} \simeq 1 - O(m_{ext}/\epsilon m_0)~.
\label{5.6}
\end{equation}
In the same vein, using Eq.(\ref{4.8}) we  get
 \begin{equation}
\gamma (R) \simeq 1 - {2m_0 \over |R|} + O(m_{ext}/\epsilon m_0)~.
\label{5.7}
 \end{equation}
Thus  $\epsilon$ and $R_A$ should be  chosen in such a
 way that $m_{ext}/\epsilon m_0 \ll 1$.
The last two expressions imply that the condition
 that backreaction is negligible is
not that $m_{ext} \ll m_0$ but rather $m_{ext} \ll \epsilon m_0$;
 in regions close to
a horizon even if   $\epsilon \ll 1$, a small cloud of matter can strongly
influence the geometry.
Estimates   (\ref{5.6}) and (\ref{5.7}) hold, for instance, for
   $R_A \ge 2m_0(1 +
\sqrt{m_{ext}/ m_0})$. In this case
\begin{equation}
1 \ge \beta(R) \ge \beta(R_A) \simeq 1 - O(\sqrt{m_{ext}/ m_0})~,
\end{equation}
and
\begin{equation}
\gamma (R) \simeq 1 - {2m_0 \over |R|} + O(\sqrt{m_{ext}/ m_0})~.
\end{equation}
In what follows our interest is focused on  the
region    $R>3m_0$, since only in that region of spacetime we
get   sensible analytic estimates. This is equivalent 
to  the choice  $\epsilon \ge 1/2$.

 For $R>3m_0$  the scalar wave equation Eq.(\ref{4.15}) can be approximated as
  \begin{equation}
(\partial_0 +\gamma \partial_R)(1 - {2m_0 \over R})h = \hat h ( 1 - {2m_0 \over R})
{m_0R \over |R|^3}~;
\label {5.8}
\end{equation}
from the preceding analysis it follows 
that the error terms are of order $m_{ext}/m_0$.
Thus, in the limit of $m_{ext}/m_0 \ll 1$ we 
recover the Schwarzschild background scalar field
equation, Eq.(\ref{3.4}), with a solution Eq.(\ref{3.7})  
consisting  both of the redshift and the
backscattering terms.

It turns out that $m_{ext}$ can be used in another role.
It is essentially the integral of the square of the 
first derivative of $\phi$ and so one can use
it to bound the pointwise value of $\phi =2\hat h$.
One can show, improving a coefficient in an 
inequality of \cite{Malec1997}, that
\begin{equation}
|\hat h|\le {\sqrt{ m_{ext}}\over R^{1/2} \sqrt{8\pi (1-{2m_0\over R})}}.
 \label {5.9}
\end{equation}
(An estimate similar to this, but written with the 
total Bondi mass $m_0$ in place of $m_{ext}$,
appears in  \cite{Demetrios}.) Thus the solution of 
Eq.(\ref {5.8}) can be estimated by solutions of
following equations
\begin{equation}
(\partial_0 + \partial_{r^*}) \Bigl( (1-{2m_0\over R})h \Bigr)
 =- { 2m_0 (1-{2m_0\over R})
\sqrt{ m_{ext}}\over R^{5/2} \sqrt{8\pi (1-{2m_0\over R})}},
 \label {5.10}
\end{equation}

\begin{equation}
(\partial_0 + \partial_{r^*})\Bigl( (1-{2m_0\over R})h \Bigr)
 = + { 2m_0 (1-{2m_0\over R})
\sqrt{m_{ext}}\over R^{5/2} \sqrt{8\pi (1-{2m_0\over R})}},
\label {5.11}
\end{equation}
where the minus and plus signs gives a bound from below and   above,
 respectively.
 The solution of Eq.  (\ref {5.10})  the same as  of  Eq.(\ref{3.7})
 \begin{equation}
\left(1-{2m_0\over R(r)}\right)h(r^* , t) = h_0(r^* - t) + 
\sqrt{{m_{ext}\over 8\pi }}2m_0
\int_{(r - t, 0)}^{(r, t)}\left(1 - 
{2m_0 \over R}\right)dv {1\over R^2 \sqrt{ R-2m_0}}
\label {5.12}
\end{equation}
and equation (\ref {5.11}) is solved by
\begin{equation}
\left(1-{2m_0\over R(r)}\right)h(r^* , t) = h_0(r^* - t) -
 \sqrt{{m_{ext}\over 8\pi }}2m_0
\int_{(r - t, 0)}^{(r, t)}\left(1 - 
{2m_0 \over R}\right)dv {1\over R^2 \sqrt{ R-2m_0}}
\label {5.13}
\end{equation}
 Notice that that the $dv$ in the integrals in Eqs.(\ref{5.12}) and
(\ref{5.13}) is essentially $dr^*$, the Regge-Wheeler
 coordinate introduced in Eq.(\ref{3.5}) with
$dR/dr^* = 1 -2m_0/R$. Therefore the integral 
can be written in closed form as
\begin{eqnarray}
\sqrt{{m_{ext} \over \pi}}m_0
\int {dR\over R^2 \sqrt{( R-2m_0}} &=
\sqrt{{m_{ext} \over 2\pi}}m_0
\left[{\sqrt{R - 2m_0}\over 2m_0R}
+ {1 \over 2m_0\sqrt{2m_0}} \arctan \sqrt{{R - 2 m_0 \over 2 m_0}}\right]
\label{5.14}\\
&= \sqrt{{m_{ext} \over 16 \pi m_0}}
\left[\sqrt{{2 m_0 \over R}\left(1 - {2m_0 \over R}\right)} +
\arctan \sqrt{{R - 2 m_0 \over 2 m_0}}\right] ~.
\label{5.15}
\end{eqnarray}

 From Eq.(\ref{5.15})  not only can we see that 
the redshift term in the solution (Eq.(\ref{5.12})
or (\ref{5.13})) is correct up to terms of order 
$m_{ext}/m_0$, we can also see that the total
backscattering is bounded by a term of order $\sqrt{m_{ext}/m_0}$.
 Thus in the limit where
$m_{ext}/m_0$ is small, the standard gravitational
 redshift represented by Eq.(\ref{3.10})  is
recovered   and this is valid along an 
outgoing null ray to an accuracy of $O\sqrt{m_{ext}/m_0}$.
Let us stress that this holds in a
 strong gravitational field. We need not assume $R \gg m_0$,  but
only $R > 3 m_0$.

The estimate of the backscattering, Eq.(\ref{5.15}), 
is amazingly regular. It equals
0 when $R = 2m_0$,    remains always positive, 
  monotonically increases and has a maximum value
of $\sqrt{\pi m_{ext}/64m_0}$ when $R \rightarrow \infty$. 
It rises quite rapidly to
 about 1/3 of its final value when $R = 3m_0$ and about 0.8
 of its final value when $R =
4m_0$. This means that the backscattering is only 
important deep in the potential well and its
contribution rapidly diminishes as one considers
 radiation which is starting off further
and further out. The redshift factor, $\gamma = 1 - 2m_0/R$ 
is much more severe
 and the energy flux (or the magnitude of $h$) 
is severely diminished if one starts close
to the horizon.

More detailed information about the nature
 and magnitude of the backscattering term will
be given in the next section.

\section{Backscattering}

Let us assume   initial data which (at $t = 0$
) represent a pure outgoing wave, i. e.,
  $h_- \equiv 0$ or  (equivalently) $h(R,0) = 0$ for $R < 0$. Let
us  take a point  to the future  $(R_1, T_1)$,    outside any horizon,
$R_1 > 2m_0$;    $R_1 \ge 3m_0$ would even be better in  
order to guarantee that bacreaction is
absent.  Consider the ingoing future radial null ray
 which passes through $(R_1, T_1)$. This will
start from the initial hypersurface at some point
$(R_0, 0)$ with $R_0 > R_1$. Along this null ray $R$ 
monotonically decreases while $m_{ext}$
monotonically increases.   Eq.(\ref{5.15})  implies
\begin{eqnarray}
\left(1 - {2m_0 \over R_1}\right)|h_-(R_1, T_1)|
 \le \sqrt{{m_{ext} \over 16 \pi m_0}}
&\left[\sqrt{{2 m_0 \over R_0}\left(1 - {2m_0 \over R_0}\right)} +
\arctan \sqrt{{R_0 - 2 m_0 \over 2 m_0}}\right. \nonumber\\ &- 
\left. \sqrt{{2 m_0 \over
R_1}\left(1 - {2m_0 \over R_1}\right)} - \arctan \sqrt{{R_1 - 2 m_0 \over 2
m_0}}\right]~,
\label{6.1}
\end{eqnarray}
 where $m_{ext}$ is $m_{ext}(R_1, T_1)$.
 This expression is valid to an accuracy of $O(m_{ext}/m_0)$ 
and has an upper bound of $\sqrt{\pi
m_{ext}/64m_0}$ which is achieved when when $T_1$ is large 
and $R_1 \approx 2m_0$. It reduces
rapidly as $R_1$ increases, it is down to 1/5 of this value 
when $R_1 = 4m_0$ as  already
discussed in Section 5. Far out along an outgoing null cone, 
where $R_1 \approx T_1
\gg m_0$ and where $R_0 \approx 2R_1$, it goes to zero like $R_1^{-3/2}$. 
In this regime   the
analysis of Sections 2 and 3 would hold.
 Therefore one could expect to approximate
$\hat h$ by $ g(R - t)/2R$.  Further, 
 approximating  $g(R - t)$ by a step function of finite
width and    substituting this into Eq.(\ref{3.7}), one   
 expects $h_-$ to fall off like
$R_1^{-3}$.

Take   an outgoing null ray through the point $(R_1, 0)$
 which forms the inner boundary of the
outgoing wave. We estimate $h_-$  along this null ray, using 
Eq.(\ref{6.1}), and $\hat h$.
 The  integration of  Eq.(\ref{4.11}) along this null ray   
  yields an estimate of
the total energy flux across this surface in the inward direction. 
This will be the total energy
loss from the outgoing wave due to backscattering.

To get explicit estimates,  let us further approximate 
the integral in Eq.(\ref{5.14}):
$R$ monotonically decreases along the ingoing lightray, so  one can
just replace the $\sqrt{1 - 2m_0/R}$ by $\sqrt{1 - 2m_0/R_1}$.  This yields

\begin{eqnarray}
\sqrt{{m_{ext} \over 2\pi}}m_0
\int_{R_1}^{R_0} {dR\over R^2 \sqrt{( R-2m_0}} &\le \sqrt{{m_{ext}R_1
\over 2\pi(R_1 - 2m_0)}}m_0 \int dR R^{-5/2}\\ &=
\sqrt{{m_{ext}R_1 \over 2\pi(R_1 -
2m_0)}}m_0\left[{2 \over 3R_1^{3/2}} -
 {2 \over 3R_0^{3/2}}\right]\label{6.3}\\
&\le\sqrt{{m_{ext}R_1 \over 2\pi(R_1 - 2m_0)}}m_0
 \left({2 \over 3}\right)R_1^{-3/2}~.
\label{6.4}
\end{eqnarray}

Thus we arrive at
\begin{equation}
|\gamma h_-(R)| \le {2\over 3} \sqrt{{m_{ext} \over 2\pi m_0}}
\left[{m_0^{3/2} \over R\sqrt{R - 2m_0}}\right]~.
\label{6.6}
\end{equation}
Knowing $ h_-$,  one can use its definition, Eq.(\ref{4.2}),
 to compute $R\phi$ and hence $\phi$ and $\hat h$.   Eq.(\ref{4.2})  yields
the inequality
\begin{equation}
(\partial_0 + \partial_{r^*})(R\hat{h}) = \gamma h_- \le {2 \over 3}
\sqrt{{m_{ext} \over 2\pi m_0}}\left[{m_0^{3/2} \over R(R - 
2m_0)^{1/2}}\right];
\label{6.7}
\end{equation}
  a similar inequality with minus sign gives a lower bound.

These can be solved in the standard way to give
 (remembering to change from $dr^*$ to $dR$)
\begin{equation}
|R\hat{h}(R,t)| \le |R\hat{h}(R_2,0)| +
{2 \over 3} \sqrt{{m_{ext} \over 2\pi m_0}}
m_0^{3/2}\int{dR \over (R - 2m_0)^{3/2}}~,
\label{6.8}
\end{equation}
where we assume that the outgoing null cone starts at $(R_2, 0)$.
  The  integration gives
\begin{equation}
|R\hat{h}(R,t)| \le |R\hat{h}(R_2,0)| +
{4 \over 3}
\sqrt{{m_{ext} \over 2\pi m_0}}{m_0^{3/2} 
\over (R_2 - 2m_0)^{1/2}} -{4 \over 3}
\sqrt{{m_{ext} \over 2\pi m_0}}{m_0^{3/2} \over (R - 2m_0)^{1/2}}~.
\label{6.8a}
\end{equation}
 Now let us choose this outgoing null ray 
inside the outgoing burst of radiation; there is
  neither an ingoing nor an outgoing field, i.e.,  
 $h_+ = h_- \equiv 0$. This does not   imply
$\phi/2 = \hat{h} \equiv 0$. Rather it only implies 
$R\phi = C$, with $C$ a constant.
 This looks like the static field of a point charge. 
If  there is  a regular center, this would be
excluded, but in the case of a black hole, or if  
there is  some sort of complicated interior
this cannot {\it a priori} be  ruled   out.
 On the other hand such a `static' field  has
energy density and this will cause the mass function
 to vary along the lightcones. We are not
interested in such variation focusing
 only in evaluating the change in the mass
function due to backscattering. 
Therefore we make the extra assumption that $\phi(R_2, 0) = 0$.
This means that the first term in Eq.(\ref{6.8a}) vanishes.

 It is clear that the last term in
Eq.(\ref{6.8a}) is strictly larger than $|h_-|$ as 
given by Eq.(\ref{6.6}) if $R > 4m_0$ (i. e.,
when {\it all} radiation is placed outside $4m_0$).  Thus
\begin{equation}
|\hat{h} - h_-| \le |\hat{h}| + |h_-| \le {4\over 3} \sqrt{{m_{ext}
\over 2 \pi m_0}}{m_0^{3/2}R_2 \over R(R_2 - 2m_0)^{1/2}} ~.
\label{6.9}
\end{equation}

 The  integration of Eq.({4.11}) from $R_A$ to $\infty$  
gives a bound for  the total
change in $m_{ext}$,
\begin{equation}
\Delta m_{ext} \le
m_{ext} {16\over 9}
\left( {2m_0 \over R_2} \right)^2
\left({1 - m_0/R_2 \over 1 - 2m_0/R_2}\right)~.
\label{6.10}
 \end{equation}
This expression demonstrates how sensitive the amount of backscattering
is to the location of the innermost null cone. 
This estimate becomes meaningless if $R_2 \approx
3.5m_0$ because   $\Delta m_{ext} \ge m_{ext}$. 
 In the case of a neutron star, where $2m_0/R \le
0.1$ (on the surface of the star) we have an upper
 bound for the backscattered energy of 2\% of
$m_{ext}$.

 The above estimation can be significantly refined 
 when initially outgoing pulses are
far enough from the apparent horizon. The 
basic point uses the following argument.
 Take an outgoing null ray which starts at a point $(R_2, 0)$
and which goes through $(R_1, T_1)$.  
Consider also the incoming null ray which starts at $(R_0,
0)$. These have corresponding tortoise coordinates
 $r^*(2), r^*(1), r^*(0)$ as defined by
Eq.(\ref{3.4}). They satisfy
\begin{equation}
r^*(0) - \tau(1) = r^*(1) = r^*(2) + \tau(1)~.
\label{6.15}
 \end{equation}
While it is NOT true that $R_0 < 2R_1$ in general, 
it is {\it almost} true and
it is certainly true, for instance when $R_2\ge 12 m_0$.
In going from Eq.(\ref{6.3}) to Eq.(\ref{6.4})  one  can
 legitimately replace $R_0$ by $2R_1$.
That allows one to get in (\ref{6.6}) the additional factor
$\alpha \equiv {\sqrt{2} -1/2\over \sqrt{2}}$. (\ref{6.6}) reads
\begin{equation}
|R\hat{h}(R,t)| \le |R\hat{h}(R_2,0)| +
2{2\sqrt{2} - 1 \over 3\sqrt{2}}
\sqrt{{m_{ext} \over \pi m_0}}m_0^{3/2}\int{dR \over (R - 2m_0)^{3/2}}~,
\label{6.16}
\end{equation}
 assuming  that the outgoing null cone starts at $(R_2, 0)$.
 While it is possible to solve this integral explicitly, 
an interesting approximation can be found
by extracting the $1 - 2m_0/R$ and replacing it with $1 - 2m_0/R_2$ to get
\begin{equation}
|R\hat{h}(R,t)| \le |R\hat{h}(R_2,0)| +
{8\sqrt{2} - 4 \over 3\sqrt{2}}
\sqrt{{m_{ext} \over 2\pi m_0}}{m_0^{3/2} \over (1 - 2m_0/R_2)^{3/2}}
\left[ {1 \over \sqrt{R_2}} - {1 \over \sqrt{R}}\right]~.
\label{6.17}
\end{equation}
Now the second term is more than twice as large as
 the estimate for $|h_-|$. Thus we may conclude
that
\begin{equation}
|\hat{h} - h_-| \le {8\sqrt{2}-4 \over 3\sqrt{2}} 
\sqrt{{m_{ext} \over 2\pi m_0}}
{m_0^{3/2} \over (1 - 2m_0/R_2)^{3/2}}
\left[ {1 \over R\sqrt{R_2}} - {1 \over 2R^{3 \over 2}}\right]~.
\label{6.18}
\end{equation}
Therefore,  the total change in $m_{ext}$   
 can be obtained  by inserting the above into
     Eq.({4.11}) and integration  from $R_2$ to $\infty$, which yields
\begin{equation}
\Delta m_{ext} \le
{\alpha^2 16 \over 9} \Bigl( {2m_0\over R_2}\Bigr)^2 
{(1 - m_0/R_2) \over 1-2m_0/ R_2}m_{ext}~.
\label{6.20}
\end{equation}
That estimate is better than the former one by
 the factor of $\alpha^2 \simeq 0.4$.
Applying that to case of neutron stars,
 when the conditions assumed above hold true, we can get
that the maximal amount of (possibly) 
backscattered radiation does not exceed 1 percent.

 The initially outgoing field, $h_+$, generates
 a weaker ingoing field, $h_-$, which enters the
`no-radiation' zone behind the wavefront. This,
 in turn, scatters again off the gravitational field
to generate a new outgoing field, which turns up at null 
infinity at a later time, after the first
burst of outgoing radiation has gone. This is called the `tail' term.
 Eq.(\ref{6.9}) gives   an
estimate  for $\hat{h}$ in the `no-radiation' zone.
 Let us substitute this back into the scalar
wave equation, say Eq.(\ref{5.8}), and estimate the second order $h_+$
 along the outgoing null ray.
 We get
\begin{equation}
|\gamma h_+(R,t)| \le {8 \over 3}
\sqrt{{m_{ext} \over 2\pi m_0}}{m_0^{5/2} \over \sqrt{R_2 - 2m_0}}
\int_{R_2}^R \left[ {1 \over R^3} - 
{\sqrt{R_2 - 2m_0}\over R^3\sqrt{R - 2m_0}}\right]dR~.
\label{6.12}
\end{equation}
Throwing  away the second term in Eq.(\ref{6.12}) and integrating
Eq.(\ref{6.12})  gives
\begin{equation}
|\gamma h_+(R,t)| \le {8\over 3}
\sqrt{{m_{ext} \over 2\pi m_0}}{m_0^{5/2} \over
\sqrt{1 - 2m_0/R_2}} \left[{1 \over 3R_2^{5/2}} - 
{1 \over 3R^2\sqrt{R_2}} \right]dR~.
\label{6.13}
\end{equation}
In the limit as $R \rightarrow \infty$ along the 
null cone we get
 \begin{equation}
|h_+(\infty,\infty)| \le {8 \over 9}
\sqrt{{m_{ext} \over 2\pi m_0}}{(m_0/R_2)^{5/2} \over \sqrt{1 - 2m_0/R_2}}~. 
\label{6.14}
\end{equation}
That should be compared with the leading term in (\ref{5.12}), $h_0$; resorting to the
definition of the external mass $m_{ext}$, we arrive to the conclusion
 that the tail term is smaller than the leading term by a factor $(m_0/R_0)^2$.

\section{Numerical results.}

In order to estimate quantitatively the effect of backscattering 
we solve numerically
Eqs.(\ref{4.16prim}, \ref{4.16} and \ref{4.4}) 
 neglecting backreaction (i.e.  dropping
terms of order $m_{ext}/m_0$) and calculate $m_{ext}(R)$ using Eq.(\ref{4.9}).

The basic evolution Eqs.(\ref{4.16prim}, \ref{4.16}) 
are solved with a help of
modified MacCormack predictor-corrector scheme \cite{hirsch}, 
\cite{seidel}. An integral defining
$\hat h$ in Eq.(\ref{4.4}) is calculated by means of the extended
 trapezoidal rule. The whole
procedure is second--order accurate in both space and time.

We use the following form of initial data:

\begin{eqnarray}
h_+(R) &= A_0 (R-R_0) exp(-20 (R-R_0)^2/s^2) \\ h_-(R) &= 0~. 
\label{num1}
\end{eqnarray}
where $A_0$ -- amplitude, $R_0$ -- position of peak center, $s$ -
the measure of its width. This data fulfills the constraint of Eq.(\ref{4.7}).

A sample of results is summarized in Figures 1 and 2 where the evolution of
 $h_+$ and $\hat h$ functions is shown.  Notice the effects of the redshift and widening of the
$h_+$ profile caused by the light-cone structure in the vicinity of the horizon.

As a quantitative measure of backscattering we take the mass flux ratio:
\begin{equation}
r_m(T)={(m_{ext}(R,T)-m_{ext}(R_i,0)) \over m_{ext}(R_i,0)}~.
\label{num2} \end{equation}
where $(R_i,0)$ and $(R,T)$ are connected by an outgoing future radial null ray.

Backscattering is noticeable   close to the horizon in the
regions where the $\hat h$ function is significant.
  We do not observe a  significant
effect outside an impulse -
 the observed flux through the inner boundary of a signal defined as a
radius which contains initially 90 \% of $m_{ext}$ is  
 less than 1 \% (Figure 3, long dashed
line). We do observe, however, a  significant
 effect due to  backscattering deep inside a region
filled with radiation. The total 
leakage of mass  is of the order of 10 \% for $R_i = R_0$, where
$R_0$ is the position of the peak center
 in the initial configuration ansatz for a configuration
starting from
$R_{0} = 1.5 R_{hor} = 3m_0$ (see Figure 4, solid line). 
This effect decreases with the distance
from the horizon, for a similar configuration starting
 from $R_0 = 3 R_{hor} = 6m_0$ the effect is
less then 2\% (see Figure 3, short dashed line).

From this we can conclude that backscattering deforms signals.
 Thus, assuming that a source mechanism for the radiation is known, 
we should in principle  be
able to detect backscattering in pulses or bursts of
 radiation coming to us from  compact objects
like massive neutron stars or black holes.

\section{Backscattering: astrophysical implications?}

 We expect that massless spin-1 field (the photons) 
will behave in a very similar fashion
to the massless spin zero field we have been discussing here. It 
is clear that two mechanisms
effect the outgoing radiation. One, the redshift effect  
diminishes the intensity and frequency of
the outgoing radiation, but the total energy in the radiation 
(as measured by the mass function)
remains unchanged. This effect becomes larger and larger as the source
 approaches the horizon; the
intensity is just multiplied essentially by $\gamma = 1 - 2m_0/R$.

The other effect is backscattering. The estimates
 that we derived in Section VI are not
  sharp but they are valid at least to within 
an order of magnitude, as the numerical results
show.   A quasi-stationary radiating plasma surrounding a black hole
  will  produce  radiation; some of the ingoing radiation 
will be reflected outwards (the black
hole can act as a kind of a mirror) 
 while some of the outgoing radiation will be reflected back.
 The reflection coefficients need not to be symmetric -
 presumably much more of the outgoing
radiation will be reflected back -   hence  the radiation 
observed at infinity should be
reduced below that expected when only the redshift 
factors are taken into account.
The overall weakening of radiation  in
 quasi-stationary systems may be   of   astrophysical
significance in modelling activity of galactic nuclei 
powered by black holes, for instance a  nuclei
 of the elliptical galaxy M87. That nuclei seems to be    much less
luminous than it should be, based on the standard accretion
 models (\cite{Rey96},   \cite{Nar}).
 Backreaction is certainly unable to  to explain the full extent of 
the existing discrepancy, but
it can contribute to  the  possible explanation if  the bulk of the 
radiation is
produced   at distances of a few Schwarzschild radii, when
 backscattering can play an important
role.

Numerical results of Section VII  demonstrate also another effect, 
namely up to $10\%$ of the scalar
field radiation emitted at $R = 3m_0$ is   lost from the main impulse -
 even if it reaches
infinity, it does so with a significant delay. More precisely,
 Figure 3 shows, that for an
extended pulse of radiation, its `first' half is weakened, while 
the other gains in intensity. Thus
the shape of the pulse is deformed. That opens another possibility
 for detecting
 backscattering, namely through  the investigation of short-lived signals 
coming from an immediate
vicinity of compact bodies.   One class of  phenomena,  X-ray bursts,
 are already
known to result from energetic processes on a surface of neutron 
stars \cite{Lewin}, so they
constitute a natural object of interest, albeit in that case the 
backscattering is rather weak, less
than 1 percent. The most  interesting situation would be   a 
 merger of a neutron star and a
black hole, since the effect would be then relatively strong. 
 If gamma ray bursts
\cite{p95} result in the
latter type of  collision, then   sufficiently precise observations 
should reveal imprints of
bacscattering in the  main burst as well in the afterglow. 
 Their definitive absence would strongly
favour alternative scenarios for the formation of gamma ray bursts.

\begin{figure}
\epsfxsize=6in
\centerline{\epsffile{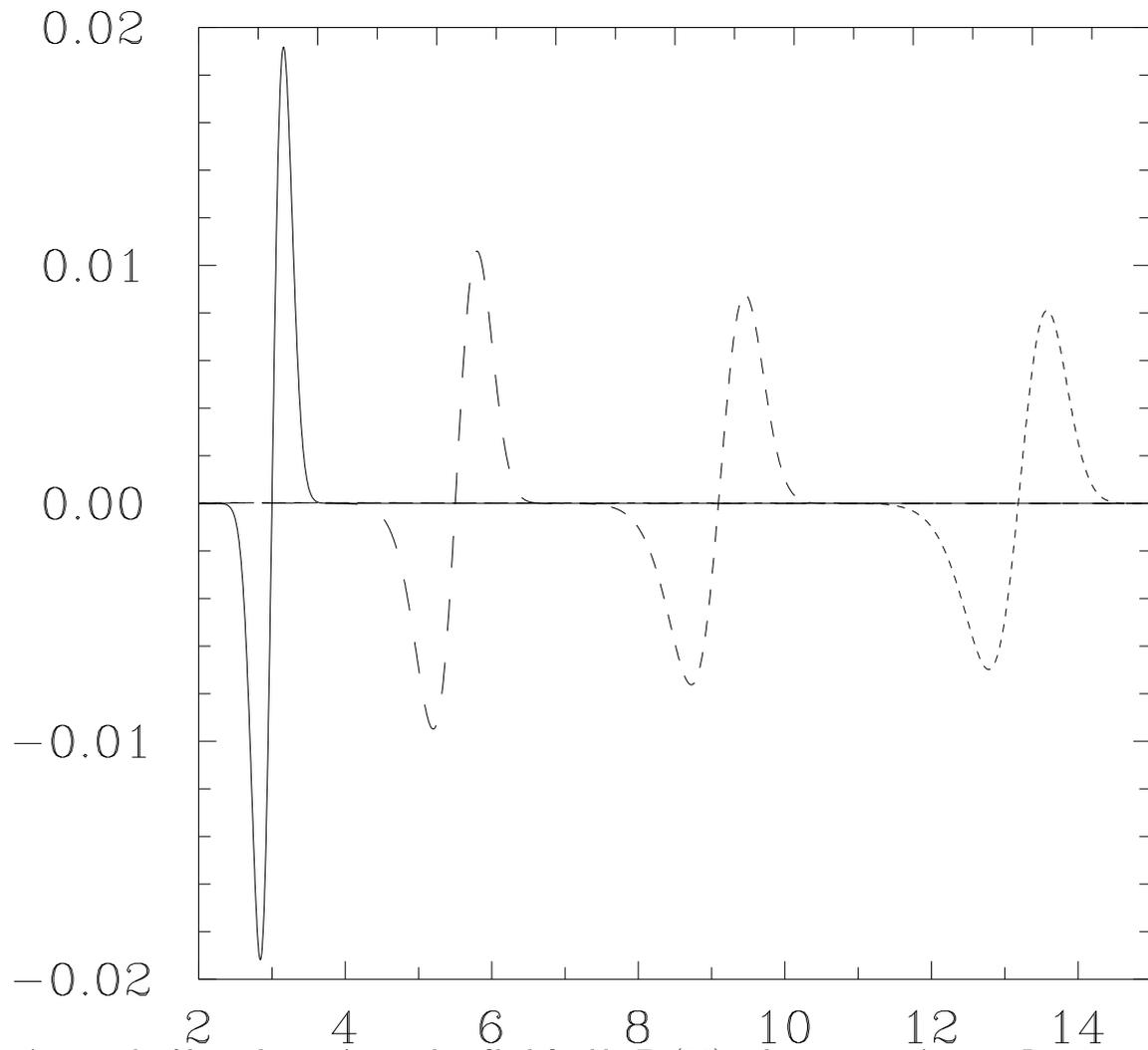}}
\caption{An example of $h_+$ evolution.
 An initial profile defined by Eq.(7.1) with parameters $A_0 = 0.2$, $R_0 = 3$, $s=1$, $m_0 = 1$ (solid line)
and profiles corresponding to $t=5$ (long dashed), $t=10$ 
(middle dashed) and $t=15$ (short dashed) are plotted.}
\label{FIG1}
\end{figure}

\begin{figure}
\epsfxsize=6in
\centerline{\epsffile{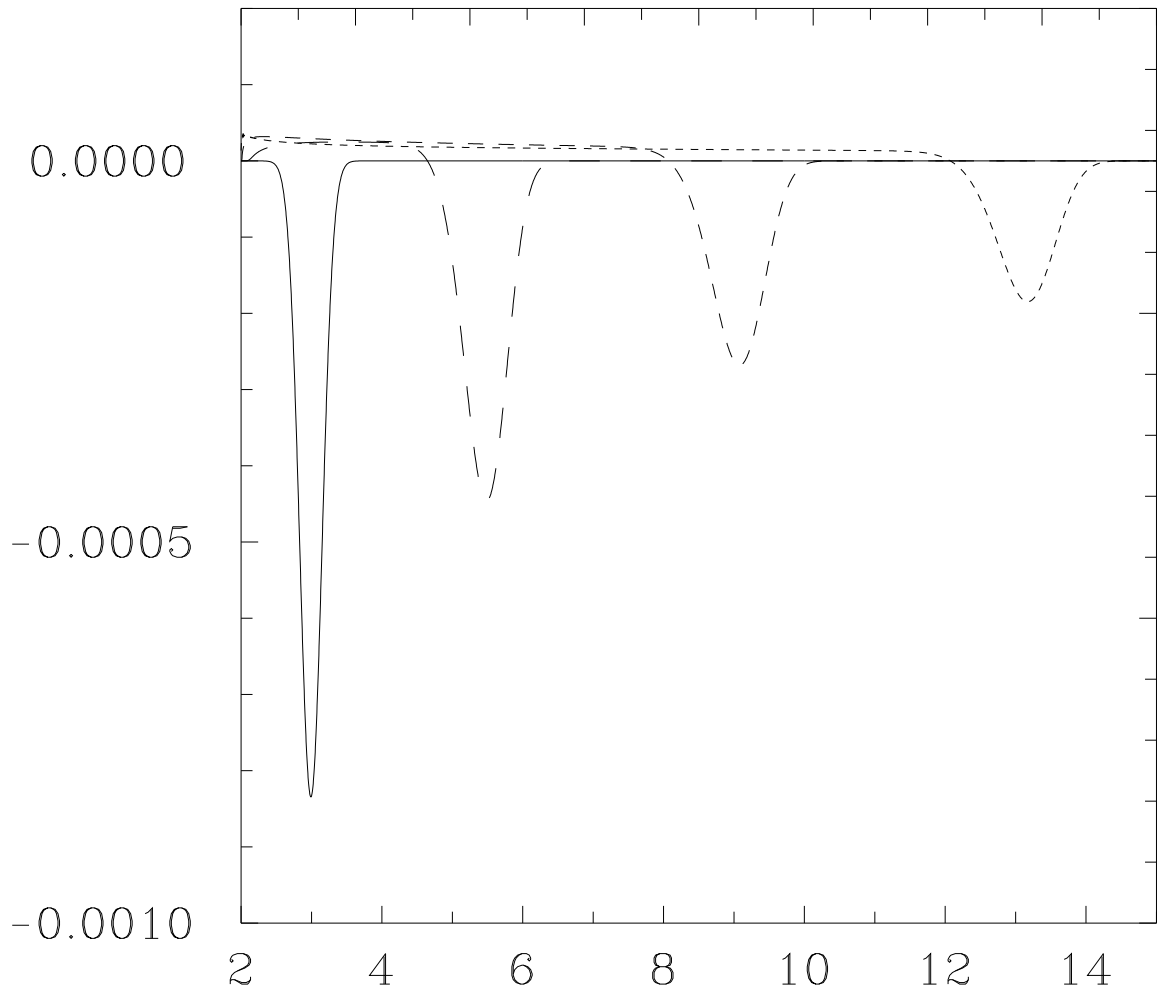}}
\caption{An example of $\hat h$ evolution. Parameters of initial confiuration
 and time sequence as
in Figure 1. }
\label{FIG2}
\end{figure}

\begin{figure}
\epsfxsize=6in
\centerline{\epsffile{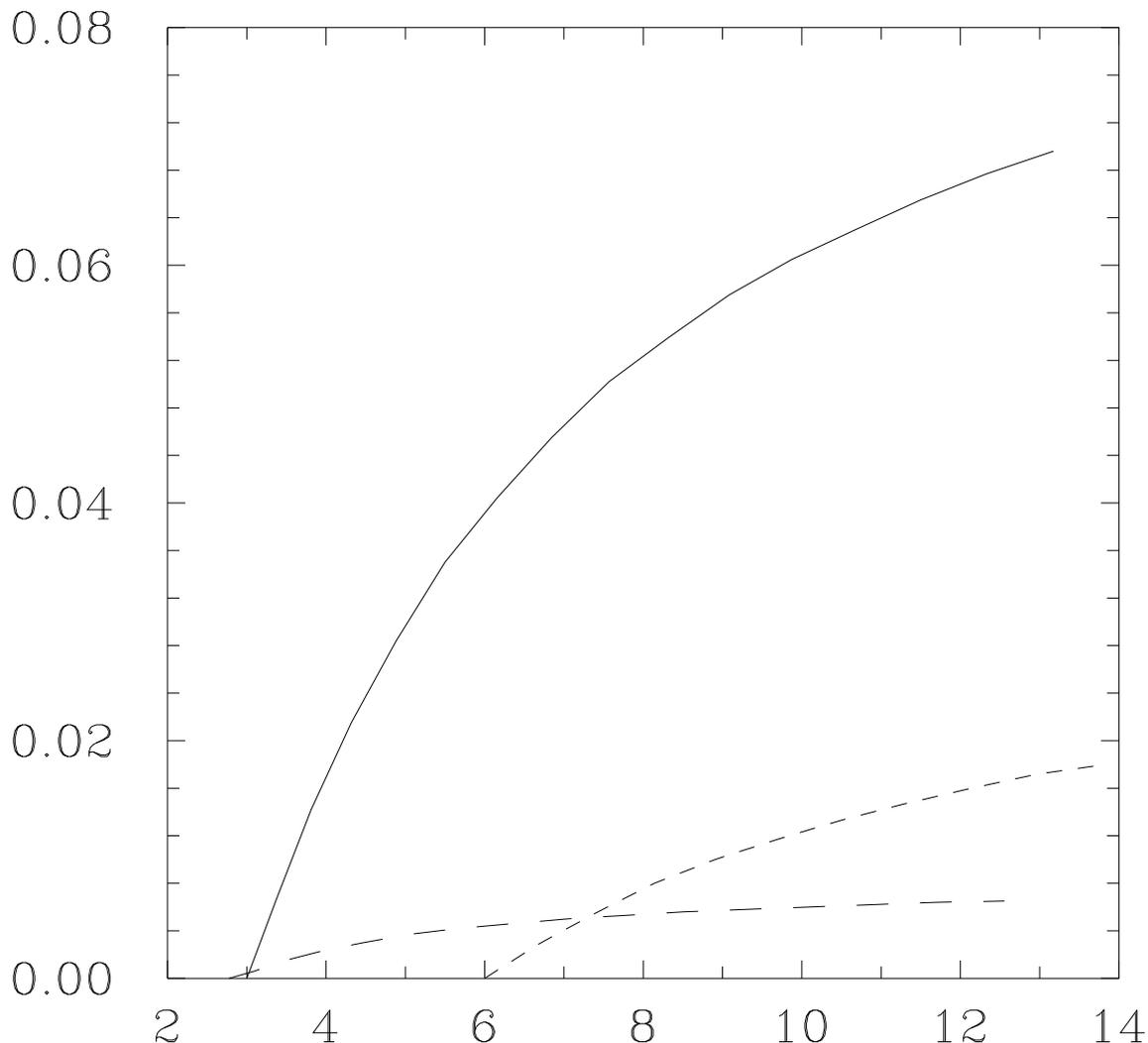}}
\caption{Examples of  leakage through null cones - 
mass flux ratio $r_m(T)$ through 
different outgoing future radial null rays as a function of position along 
the ray $R(T)$. The curves correspond to the following sets of parameters: 
solid line -- $R_i=R_0=3, s=1, A_0=0.2, m_0=1$; short dashed 
line -- $R_i = R_0 = 6, s=1, A_0=0.2, m_0=1$; long dashed line --
 $R_i = 2.776, s=1, A_0=0.2, m_0=1$.} \label{FIG3}
\end{figure}

\vskip 1cm
\acknowledgements This work has been partially supported by the Forbairt grant SC/96/750 and
the KBN grant 2 PO{3}B 090 08.


\begin{references}
\bibitem{bachelot} A. Bachelot, {\it J. Math. Pures Appl.} {76}, 155(1997).
N. Sanchez,{\it J. Math. Phys.} {\bf 17}, 688(1976).
\bibitem{Wald} R. Wald, {\it General Relativity}, Chicago University Press 1984.
\bibitem{MTW} C. Misner, K. Thorne, J. A. Wheeler, {\it Gravitation}, Freeman,
San Francisco,1973.
\bibitem{Malec1996b} E. Malec {\it Classical and Quantum Gravity}
 {\bf 13}, 1849(1996).
\bibitem{Malec1997} E. Malec, {\it Journal of Mathematical Physics}
 {\bf 38}, 3650(1997).
\bibitem{Demetrios} D. Christodoulou, {\it Commun. Math. Phys.} {\bf 106},
487(1987).
\bibitem{Rey96} C. S. Reynolds et al., {\it The "quiescent"
 black hole in M87}, astro-ph/9610097.
\bibitem{Nar} R. Narayan and I. Yi, {\it Astrophysical Journal} 
{\bf 452}, 710(1995);
Abramowicz et al., {\it Astrophysical Journal} {\bf 438}, L37(1995).
\bibitem{hirsch} C. Hirsch, {\it Numerical computation of internal and
external flows}, Vol.2 (Wiley, New York, 1990).
\bibitem{seidel} E. Seidel and W.M. Suen, {\it Phys. Rev.} {\bf D42},
 384 (1990).
\bibitem{Lewin} W. H. Lewin, J. von Paradijs and R. Tamm, in: X-Ray Binaries
 (Eds: G. Lewis,
J. von Paradijs and E. van den Heuwel), Cambridge University Press (1995).
\bibitem{p95} for a review see B. Paczy\'nski, {\it Publ. astron. Soc. Pac.} 
{\bf 167},
1167(1995).
\end{references}
\end{document}